\documentclass[]{interact}

\usepackage{epstopdf}
\usepackage{subfigure}
\usepackage{color}

\usepackage[numbers,sort&compress,merge]{natbib}
\bibpunct[, ]{[}{]}{,}{n}{,}{,}
\renewcommand\bibfont{\fontsize{10}{12}\selectfont}

\theoremstyle{plain}

\theoremstyle{definition}

\theoremstyle{remark}

\begin{document}
\title{Computer simulation study of the nematic-vapour interface in the Gay-Berne model}
\author{\name{Luis F. Rull\textsuperscript{a} and Jos\'e Manuel Romero-Enrique\textsuperscript{a}
\thanks{CONTACT Jos\'e~M. Romero-Enrique. Email: enrome@us.es}}
\affil{
\textsuperscript{a} Departamento de F\'{\i}sica At\'omica, Molecular y Nuclear,
Area de F\'{\i}sica Te\'orica, Universidad de Sevilla, Avenida de Reina 
Mercedes s/n, 41012 Sevilla, Spain}}
\maketitle
\begin{abstract}
We present computer simulations of the vapour-nematic interface of the 
Gay-Berne model. We considered situations which correspond to either prolate
or oblate molecules. We determine the anchoring of the nematic phase and correlate
it with the intermolecular potential parameters. On the other hand, we evaluate the
surface tension associated to this interface. We find a corresponding states law for 
the surface tension dependence on the temperature, valid for both prolate and oblate molecules.  
\end{abstract}
\begin{keywords}
Monte Carlo simulation; liquid crystal; Gay-Berne model; interfaces; anchoring; surface tension
\end{keywords}


\section{Introduction}

Liquid crystals are states of the matter which present long-ranged orientational order \cite{deGennes}. However, the 
presence of boundaries and/or external fields can frustrate this order, leading to the appearance of elastic
deformations on the orientational order, or the emergence of topological defects. On the other hand, the bounding
surface can also control the orientational ordering. In particular, the interfacial behaviour of liquid crystals shows 
peculiarities not observed in simple fluids, as the far-field orientational field can be determined by the presence of the
interface. The nematic director orientation on both free nematic interfaces (i.e. in equilibrium
with its vapour above the nematic-isotropic-vapour triple point) and isotropic-nematic interfaces
can be either parallel, perpendicular or oblique with respect to the interface normal \cite{Sluckin}, 
leading to the so-called homeotropic, planar and tilted anchoring, respectively. The interfacial properties
of the liquid crystals are essential for the development of display technologies \cite{deGennes,Kleman}. On the other hand,
the nematic anchoring in the nematic-vapour
interface, together with the elastic properties of the liquid crystal, are determinant to understand the conformation of 
nematic droplets \cite{rull,zannoni,zannoni2}, which have applications in molecular sensing 
\cite{lin,tomar,bera,khan} or 
electro-optic devices such as privacy windows \cite{sutherland1,sutherland2,bowley,jazbinsek,alberto1,alberto2}.
From a microscopic point of view, the 
interfacial anchoring is well known to be determined by the intermolecular potential details.
A generalized Van der Waals theory \cite{Margarida}, based in the expansion of the intermolecular 
potential in spherical harmonics, shows that the three types of anchoring can be observed 
with the appropriate choice of parameters \cite{mia3,mia4}. On the other hand, molecular simulations
have also been used to understand how the interfacial anchoring is controlled by the interaction potential.
For example, for hard body potentials it is shown that the anchoring is planar for prolate molecules and
homeotropic for oblates \cite{Margarida}. When attractive interactions are included, its anisotropy also 
plays a role in the anchoring. For example, studies on lattice models show that different choices for the
parameters which define the attractive part of the intermolecular potential can lead to either homeotropic
or planar anchoring \cite{Bates1}. 

In this Paper we wish to consider the effect of both the repulsive and attractive parts of the intermolecular
potential on the interfacial anchoring of nematic liquid crystals in equilibrium with their vapours, as well as the
associated surface tension. For this 
purpose, we consider the Gay-Berne model \cite{Gay-Berne}, which has been extensively studied in the literature
to model realistic liquid crystals. This model is characterized by two parameters $\kappa$ and $\kappa'$ 
which determine the anisotropy of the repulsive and attractive parts of the intermolecular potential, 
respectively, together with two adjustable constants $\mu$ and $\nu$ (see next Section). 
For example $(\kappa,\kappa',\mu,\nu)=
(4.4,20,1,1)$ has been used to parametrise the prolate $p$-terphenyl molecule \cite{Luckhurst,Bates5}, and 
$(0.345, 0.2, 1, 2)$ for the triphenylene core discotic model \cite{Emerson2}. Note that the parametrisation can change
drastically when considering different mesogens, so it would be useful to identify which potential parameters are 
determinant for the liquid crystal interfacial properties.
There are only a few computer simulation studies on interfaces of the Gay-Berne model, in comparison with
other models, such as hard rod fluids \cite{vanRoij}. Regarding the nematic-vapour
interface, there are reports on prolate molecules with $\kappa=2$ and $3$ which show the possibility of having either homeotropic or
planar anchoring \cite{Elvira,Emerson1,Mills}. These results could be explained by heuristic arguments which identify
the ratio $\kappa/\kappa'$ as the key factor to determine the nematic phase anchoring. On the other hand, in the nematic-isotropic
interface, only planar anchoring has been reported \cite{Bates2,Bates3}. Finally, we are not aware of any
study which considers the nematic-vapour coexistence for oblate molecules. 
Discotic liquid crystals stand for promising
materials for technological applications \cite{bushby} and they have been recently studied by using Gay-Berne 
\cite{cienega,sidky} and related models such as the Gay-Berne-Kihara model \cite{bruno1,bruno2}.

The main goal of this paper is to study the nematic-vapour
interface for a wide range of intermolecular potential parameters, corresponding to both prolate and oblate 
molecules, in order to identify the key elements of the intermolecular potential which determine the 
anchoring of the nematic with respect to its interface with a vapour phase. The paper is organized as follows. In Section \ref{sec2} we will present the model and the simulation details. Our simulation results will be discussed in Section \ref{sec3}. We will end up with our conclusions in Section \ref{sec4}.    

\section{Simulation details \label{sec2}}

The Gay-Berne intermolecular potential between the molecules $i$ and $j$ can be written as \cite{Gay-Berne}:
\begin{equation}
U_{ij}^{GB} ({\bf r}_{ij},{\bf u}_i,{\bf u}_j)
= 4\varepsilon (\hat{\bf r}_{ij}, {\bf u}_i, {\bf u}_j)
\left[\rho_{ij}^{-12}-\rho_{ij}^{-6}\right]
\label{GB}
\end{equation}
where
\begin{equation}
\rho_{ij} = \frac{ r_{ij} - \sigma (\hat{\bf r}_{ij}, 
{\bf u}_i, {\bf u}_j) + \sigma_0 }{\sigma_0}
\end{equation}
$\bf u_{i}$ is the unit vector along the symmetry axis of the molecule $i$,
$r_{ij}=|{\bf r}_i - {\bf r}_j|$ is the distance along the intermolecular 
vector ${\bf r}_{ij}$ joining the centers of mass of the molecules and
$\hat{\bf r}_{ij}={\bf r}_{ij}/r_{ij}$. The anisotropic 
contact distance $\sigma (\hat{\bf r}_{ij}, {\bf u}_i, {\bf u}_j)$ and the 
depth of the interaction energy $\varepsilon (\hat{\bf r}_{ij}, {\bf u}_i, 
{\bf u}_j)$ depend on the orientational unit vector, the length to breath 
ratio ($\kappa = \sigma_{ee}/\sigma_{ss}$) and the energy depth anisotropy
($\kappa'= \epsilon_{ss}/\epsilon_{ee}$), which are defined as the ratio
of the size and energy interactions parameters in the end-to-end ($ee$) and 
side-by-side ($ss$) configurations. Their expressions are given in terms
of the lengthscale $\sigma_0$ and the energy unit $\epsilon_0$ as:
\begin{eqnarray}
\frac{\sigma (\hat{\bf r}_{ij}, {\bf u}_i, {\bf u}_j)}{\sigma_0}&=&
\Bigg[1-\frac{\chi}{2}\Bigg(\frac{(\hat{\bf r}_{ij}\cdot {\bf u}_i
+\hat{\bf r}_{ij}\cdot {\bf u}_j)^2}{1+\chi({\bf u}_i \cdot {\bf u}_j)}
+\frac{(\hat{\bf r}_{ij}\cdot {\bf u}_i
-\hat{\bf r}_{ij}\cdot {\bf u}_j)^2}{1-\chi({\bf u}_i \cdot {\bf u}_j)}
\Bigg)\Bigg]^{-1/2}\\
\frac{\varepsilon (\hat{\bf r}_{ij}, {\bf u}_i,{\bf u}_j)}{\epsilon_0}&=&
[\epsilon_1({\bf u}_i,{\bf u}_j)]^{\nu}\times [\epsilon_2(\hat{\bf r}_{ij},
{\bf u}_i,{\bf u}_j)]^{\mu}
\end{eqnarray}
where 
\begin{eqnarray}
\epsilon_1({\bf u}_i,{\bf u}_j)&=&
[1-\chi^2({\bf u}_i \cdot {\bf u}_j)^2]^{-1/2}
\\
\epsilon_2(\hat{\bf r}_{ij},{\bf u}_i,{\bf u}_j)&=&
1-\frac{\chi'}{2}\Bigg[\frac{(\hat{\bf r}_{ij}\cdot {\bf u}_i
+\hat{\bf r}_{ij}\cdot {\bf u}_j)^2}{1+\chi'({\bf u}_i \cdot {\bf u}_j)}
+\frac{(\hat{\bf r}_{ij}\cdot {\bf u}_i
-\hat{\bf r}_{ij}\cdot {\bf u}_j)^2}{1-\chi'({\bf u}_i \cdot {\bf u}_j)}
\Bigg]
\end{eqnarray}
$\chi=(\kappa^2-1)/(\kappa^2+1)$ and $\chi'=[(\kappa')^{1/\mu}-1]/
[(\kappa')^{1/\mu}+1]$. As in the original paper we choose $\mu=2$ and $\nu=1$ \cite{Gay-Berne}. 
The parameter $\kappa$ characterizes the molecular elongation along the main symmetry axis of the molecule.
So, values of $\kappa>1$ correspond to prolate molecules, and $\kappa < 1$ to oblates.
On the other hand, if $\kappa=\kappa'=1$, the Lennard-Jones potential is recovered.

In order to study the interfacial properties of the Gay-Berne model, we need to know its phase diagram.
Unfortunately there are only a few computer simulation studies which consider the dependence of
the phase diagram on the potential parameters. For $\kappa=3$, the effect of $\kappa'$ on the liquid-vapour
coexistence was studied in Ref. \cite{deMiguel1} by Monte Carlo simulations in the Gibbs ensemble and 
using Gibbs-Duhem techniques. It was found evidences of nematic-vapour and isotropic liquid-vapour
coexistence for $\kappa'=1$ and $\kappa' = 1.25$. 
On the other hand, by fixing the value of $\kappa'=5$, which was considered previously for $\kappa=3$ \cite{mia1}, 
the authors considered the effect of increasing the
molecular elongation from $\kappa=3.2$ to $\kappa=4$ \cite{Brown}, finding that for $\kappa=4$ the 
vapour-isotropic critical point disappeared and only liquid isotropic-nematic coexistence is found.
Regarding oblate molecules, most of the simulation studies focus on the nematic-columnar phase transition
\cite{Emerson2,Bates4,Ryckaert1,Ryckaert2,Chakrabarti}. However, no systematic study on the effect 
of $\kappa$ and $\kappa'$ has been performed for oblate molecules. 

Due to the limited number of studies reported in the literature which consider the nematic-vapour coexistence
in the Gay-Berne model, and the high computational time cost to obtain the full phase diagram, 
instead we followed the procedure used in Ref. \cite{rull} to generate configurations of the liquid 
(either nematic or isotropic) near coexistence with the corresponding vapour phase. 
In this technique, $NpT$ Monte Carlo simulations are performed at $p=0$ from initial conditions such that
the density is higher than the corresponding to the liquid in coexistence with the vapour at the same
temperature. Due to the anisotropic character of the molecules, the volume changes were done by either changing isotropically
the three box side lengths, keeping the box shape cubic, or by selecting randomly a side length to be changed. If the number 
of particles is large enough, the bulk results obtained in both procedures is the same. 
The simulations were organized in cycles, each one corresponding to $N$ attempts to move and rotate a randomly chosen 
particle, and a volume change try, with $N$ being the number of particles in the simulation box. 
During the simulation, the system density decreases until it reaches a value close to the 
coexistence value. If the densities of the coexisting phases are very different,  large volume change
fluctuations are rare enough to prevent the system to go to the vapour phase for short simulations. 
On the other hand, the values of the maximum attempted volume change in the simulation, $\Delta V$, are
small but such that the acceptance ration for the volume change in the simulation is always smaller
than $35\%$.  

Once we get a typical near-coexistence liquid configuration, whe put two empty boxes aside the original and
perform $NVT$ simulations in this new box using periodic boundary conditions on the enlarged simulation box. 
Now a simulation cycle corresponds to $N$ attempts to move and rotate a randomly chosen particle.
With this procedure, two (nematic or isotropic)
liquid-vapour interfaces are formed perpendicular to the longest simulation side direction, which we set to be parallel to
the $z$ axis, provided that the temperature is below the critical point value.  
Now we increase or decrease the temperature until the nematic-isotropic-vapour
or the vapour-nematic-smectic(columnar) triple point is reached. For these
simulations, we use as initial configuration the output of the simulation from the previous temperature.

Instead the full Gay-Berne potential Eq. (\ref{GB}) we used a truncated and unshifted
version which is zero when the
distance between the molecular centers of mass is larger than some cutoff value $r_c$.
No long-range corrections are considered.
With respect to this value, there is some controversy regarding its choice. For example, in Ref. \cite{Gloor}
a $30\%$ increase in the surface tension of a Lennard-Jones fluid was reported when the cutoff changed from
$r_c=2.5\sigma$ to $3\sigma$. In order to assess the optimal value of $r_c$, we performed simulations for
different values of $r_c$ for the case reported in Ref. \cite{Elvira}. Our results are shown in Table \ref{table1},
indicating that, for $\kappa=3$, we do not find significative differences for $r_c$ larger than $5\sigma_0$.
So, we choose $r_c=(\kappa+2.0) \sigma_0$ for prolate molecules, while for oblates we will consider
a fixed value $r_c=3\sigma_0$. Nevertheless, long-range corrections have been shown to be important
for the evaluation of the surface tensions in Lennard-Jones-based systems \cite{Felipe,Felipe2,Felipe3}.

During the $NpT$ simulation, we monitor bulk quantities like density $\rho_b=N/\langle V \rangle$, 
with $N$ and $V$ being the number of particles and the simulation box volume, respectively; the
nematic order parameter $S$, defined as the largest eigenvalue of the nematic order tensor 
\begin{equation}
Q  = \left\langle \frac{1}{N}\sum_{i=1}^N \frac{3{\bf u}_i\otimes{\bf u}_i-I}{2}
\right\rangle 
\end{equation}
with the associated eigenvector is the nematic director; or the average potential energy $\langle \sum_{i<j} U^{GB}_{ij}\rangle$. 
In addition, during the $NVT$ simulations we evaluate the density profile as:
\begin{equation}
\rho (z) = \frac {1}{V_b} \left\langle \sum_{i=1}^{N} \Theta (z_i-z)\Theta(z+\Delta z-z_i)\right\rangle
\end{equation}
where $V_b=L_x L_y\Delta z$ is the volume of a slice perpendicular to the $z$ axis, and $\Theta(x)$ is the Heaviside 
step function. 
We choose $\Delta z=0.15\sigma_0$. The orientational order profile is characterized by the orientational order profile:
\begin{equation}
\overline{P}_2(z) = \left\langle \frac {1}{2N_k} \sum_{i=1}^{N_k} (3 u_{z,i}^2-1)\right\rangle
\end{equation}
$N_k$ is the number of particles on the slice $k$, and $u_{z,i}$ is the $z$-component of the $i$-particle orientation unit vector $\mathbf{u}_i$. 
If particles are perfectly aligned parallel to the interface (i.e. perfect planar anchoring), $\overline{P}_2=-1/2$. On the other hand,
if molecules are pefectly aligned along the normal to the interface (i.e. perfect homeotropic anchoring), $\overline{P}_2=1$.

Finally, the surface tension $\gamma$ was evaluated by using two different methods. First, we evaluated it by using the
virial route:
\begin{equation}
\gamma_{virial} = \frac{1}{2} \int_0^{L_z} d z (P_N(z)-P_T(z))
\end{equation} 
where the $1/2$ factor is due to the presence of two liquid-vapour interfaces, and $P_N(z)$ and $P_T(z)$ are the normal 
and tangential components of the pressure tensor profile evaluated along the $z$ axis:
\begin{equation}
P_N(z) = \rho(z)k_B T-\frac {1}{2V_b} \left\langle \sum_{i,j}^{(k)} z_{ij}
 \frac {\partial {U_{ij}}}{\partial {z_{ij}}} \right\rangle
\end{equation}
\begin{equation}
P_T(z) = \rho(z)k_B T-\frac {1}{2V_b} \left\langle \sum_{i,j}^{(k)} \frac{1}{2}\left(x_{ij}
 \frac {\partial {U_{ij}}}{\partial {x_{ij}}}+ y_{ij}
 \frac {\partial {U_{ij}}}{\partial {y_{ij}}}\right)\right\rangle =\frac{P_x(z)+P_y(z)}{2}
\end{equation}
where the superindex $k$ refers to the pair of molecules $i$ and $j$ where at least one of them is on the $k$ slice.

One of the main issues for the surface tension evaluation by computer simulation is the choice of the number of particles.
For hard body simulations under similar conditions, McDonald {\it et al.} found unexpected differences between the 
$P_x$ and $P_y$ components of the pressure tensor \cite{Allen}. In order to avoid these finite size we compared
our results with those obtained by using the expanded ensemble technique \cite{deMiguel2}. This method allows an efficient calculation of the free energy difference between two interfacial states at the same temperature, number of 
particles and volume, but different interfacial area.
During the simulation, we combine translational and orientational moves with
changes of the interfacial area.
As the simulations are performed keeping $V$ constant, an interfacial area change $\Delta A$ corresponds to 
a rescaling of the center of mass positions of the particles. The acceptance
probability of transition between two states with different
interfacial areas $A$ and $A+\Delta A$ in the expanded ensemble is: 
\begin{equation}
P_{ij} = \min \{1,\exp[-\beta (\Delta U_{ij}- \Delta W_{ij}]\}  
\end{equation}     
where $\Delta U_{ij}$ is the change of potential energy and $\Delta W_{ij}$ is the difference between the
weight factors associated to the states with different interfacial area. 
We use two possible states: the unchanged state $0$, and a state where the
interfacial area has changed by a value $\Delta A$.
This procedure allows to evaluate the free-energy change $\Delta F$
as $\Delta F= -k_B T \ln (p_1/p_0) + \Delta W$, with $p_1$ and $p_0$ being the probabilities to visit
the states $0$ y $1$ during the simulations in the expanded ensemble simulation
(for more details see Ref. \cite{deMiguel2}).  
Finally, the surface tension can be evaluated as 
\begin{equation}
\gamma_{EE} = \frac {\Delta F} {\Delta A}
\end{equation}

Table \ref{table2} shows the values of the surface tension using the expanded ensemble technique for $\kappa=6$,
which is the most sensible case to the finite-size effects. Three different options to 
change the interfacial area were used: by changing only the length along either the $x$ or the $y$ axis, or by
changing both $L_x$ and $L_y$ simultaneously. The probability $p_1$ to find any of the two states is also reported.
As it can be seen, $N=4000$ is the better choice as the differences between the different estimates are within the
statistical errors.  

For the simulation, we use reduced units. So, the reduced values of density $\rho$, energy $U$, temperature $T$ and 
surface tension $\gamma$ are defined as
\begin{equation}
\rho^*=\rho\sigma_0^3\qquad U^*=\frac{U}{\epsilon_0} \qquad T^*=\frac{k_BT}{\epsilon_0}\qquad \gamma^*=\frac{\gamma \sigma_0^2}{\epsilon_0}
\end{equation}

\section{Results and discussion \label{sec3}}

\begin{figure}
\begin{tabular}{cc}
\includegraphics[width=0.45\columnwidth]{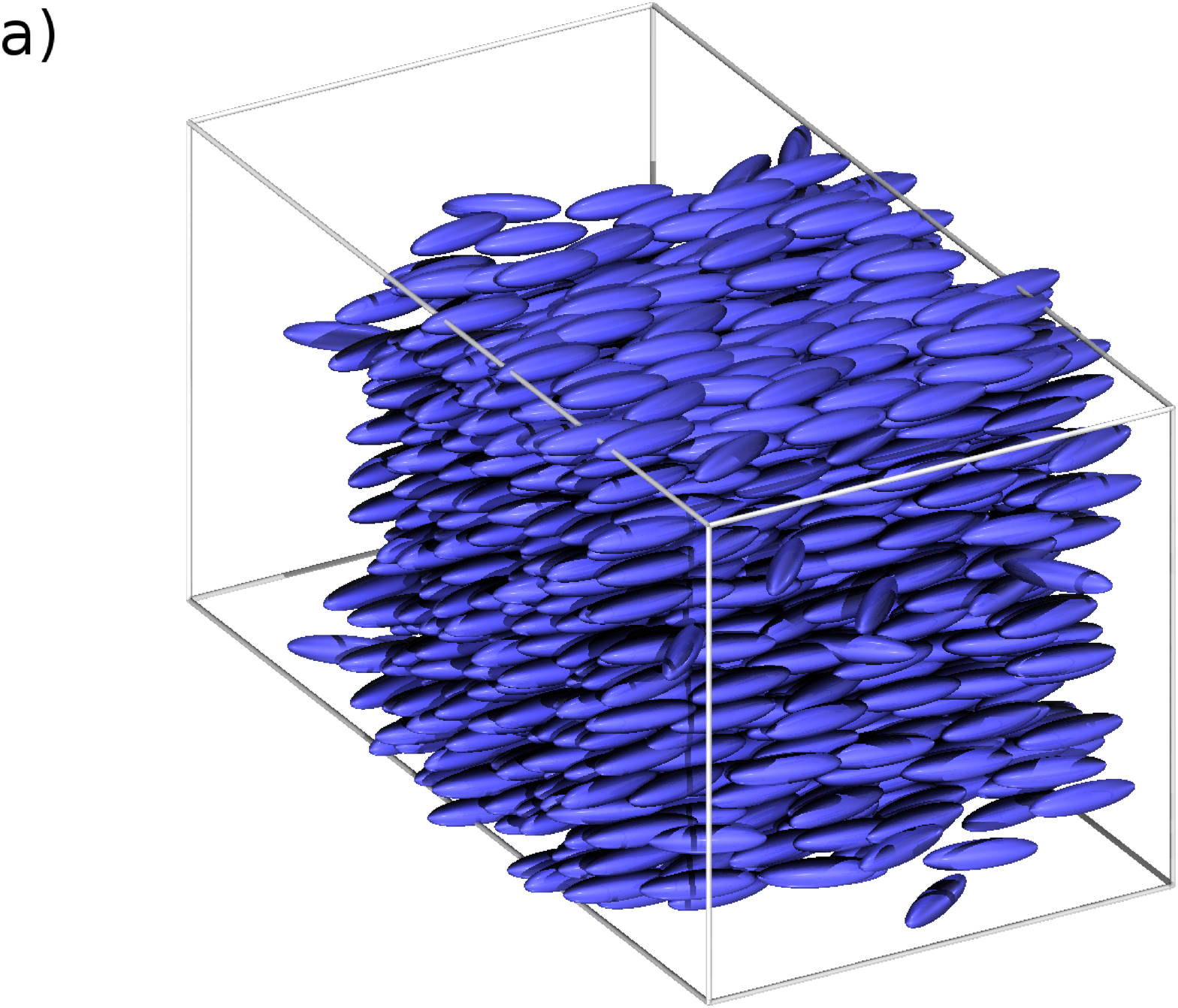} &
\includegraphics[width=0.45\columnwidth]{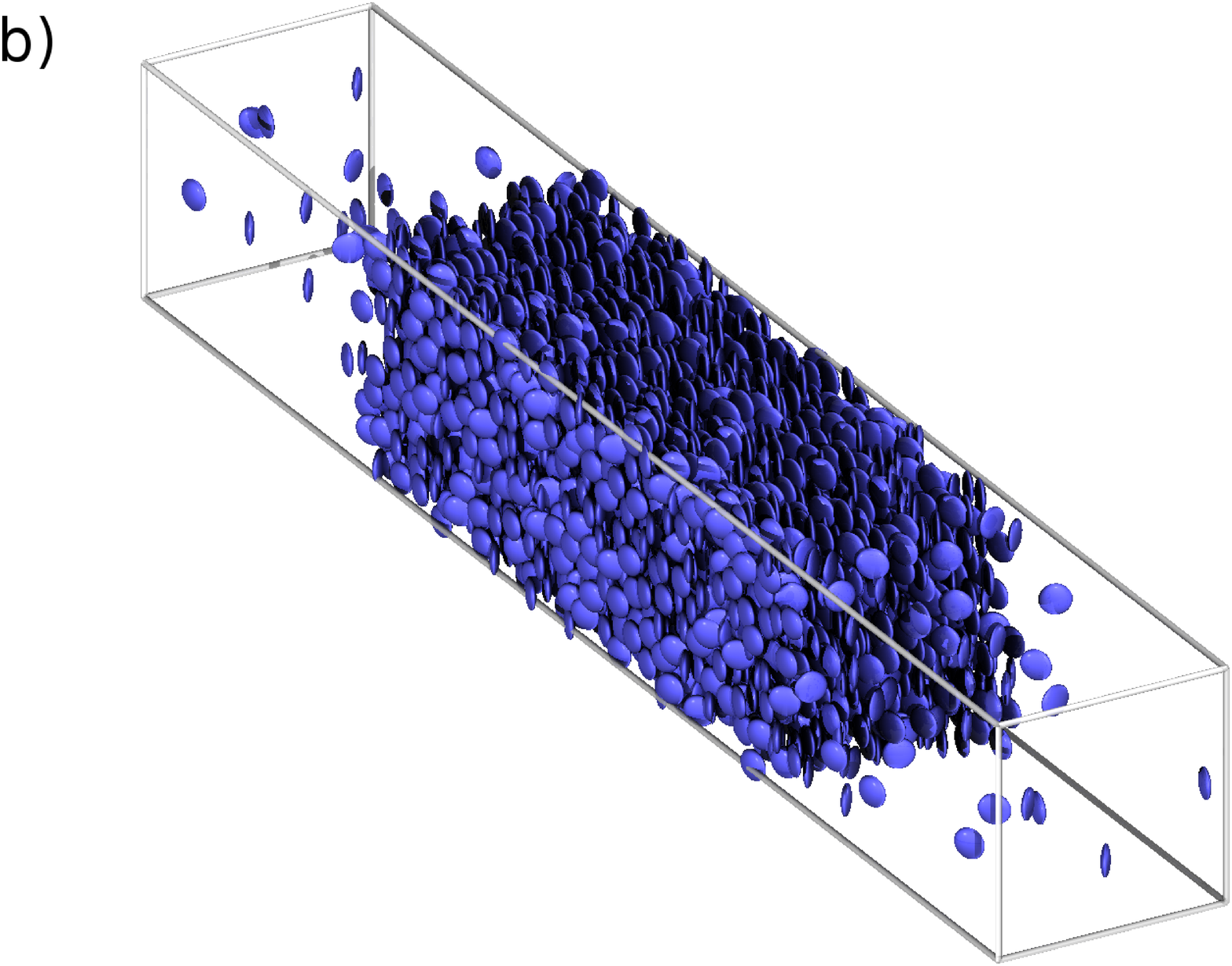}\\
\includegraphics[width=0.45\columnwidth]{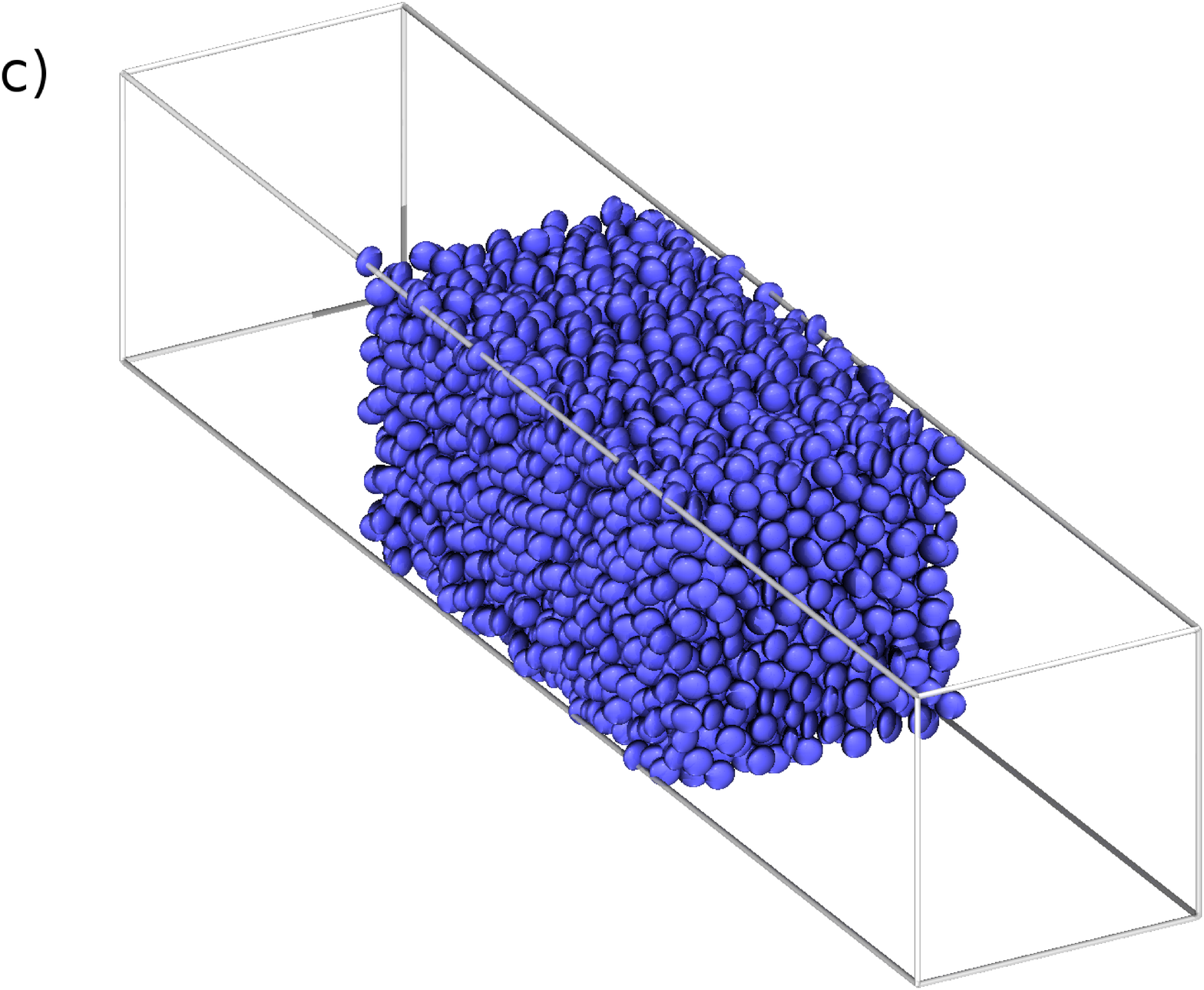} &
\includegraphics[width=0.45\columnwidth]{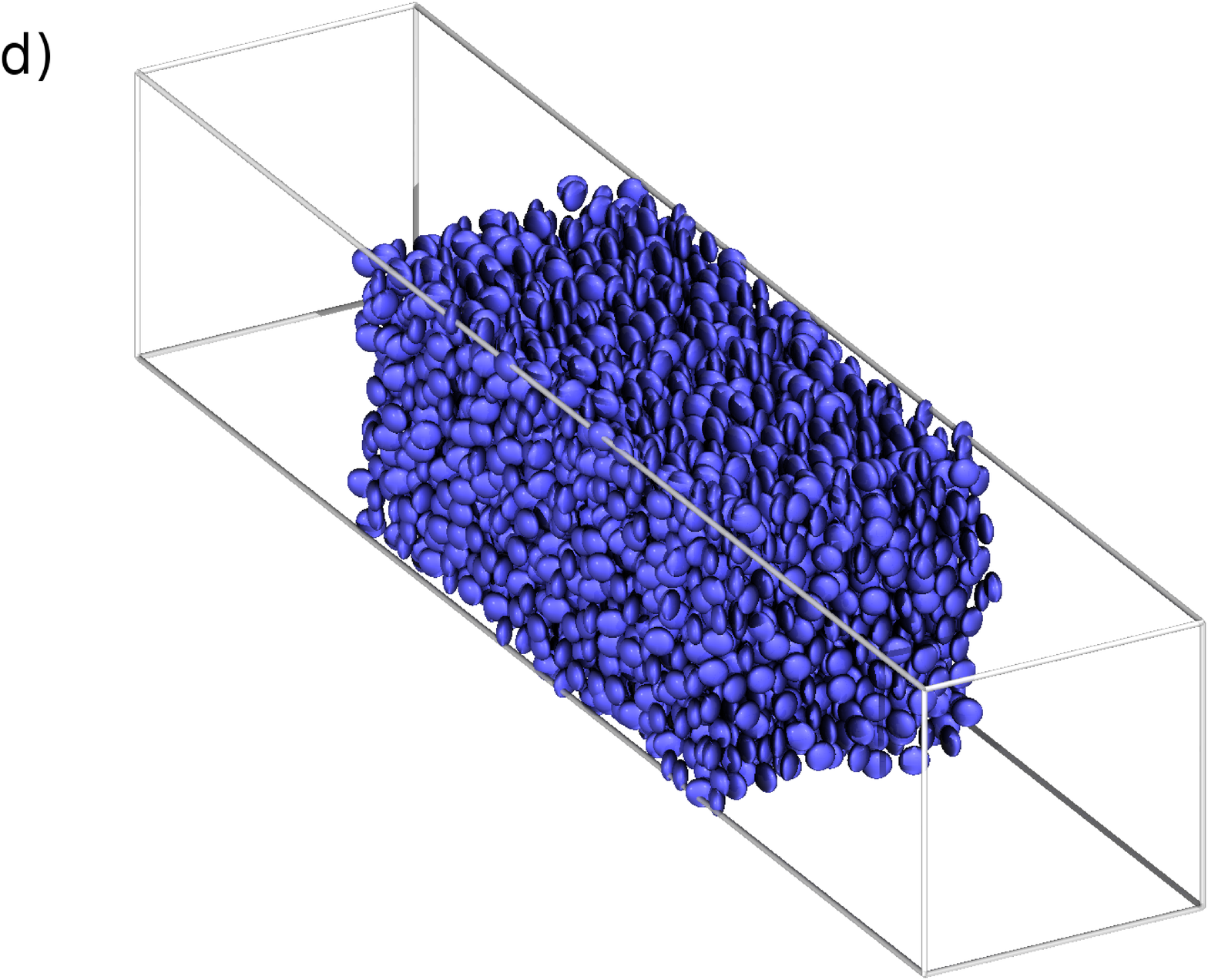}\\
\end{tabular} 
\caption
{Snapshots of the interfacial simulations of the GB model with: (a) $\kappa=4$, $\kappa'=0.5$, $T^*=1.20$ 
and $N=1372$; (b) $\kappa=0.3$, $\kappa'=0.2$, $T^*=2.50$ and $N=4116$; (c) $\kappa=0.5$, $\kappa'=0.6$, $T^*=0.40$ and 
$N=4116$; and (d) $\kappa=0.5$, $\kappa'=1.0$, $T^*=0.55$ and $N=4116$.
\label{fig0}}
\end{figure}

We have studied by computer simulations the nematic-vapour interface for different Gay-Berne models, 
which mimic either calamitic or oblate nematogens. Typical snapshots of our simulations are shown in Fig. \ref{fig0}. 
For prolate molecules, we have studied the values of $\kappa=4$ and $6$, and for oblate molecules we have considered 
$\kappa=0.3$ and $0.5$. Different values of $\kappa'$ are analysed for each $\kappa$. 
For prolate molecules, we followed the strategy considered in Ref. \cite{Brown}. For $\kappa=4$ we started with $\kappa'=1$,
which corresponds to a situation where nematic-vapour coexistence was found for $\kappa=3$, and we decreased the
value of $\kappa'$ searching for this coexistence.  
Table \ref{table3} shows the results obtained for prolate molecules. The value of $\rho_{bulk}^*$ corresponds to the
value obtained for the liquid phase in the $NpT$ simulations, except for the
cases where smectic phases appear, which were not observed in the $p=0$ 
simulations. The values of $\rho_{bulk}^*$ are almost 
identical to those obtained
from the density profile in the $NVT$ simulations well inside the liquid region. In addition, we report the type 
of phase (either nematic phase $N$ or smectic phase $Sm$), as well as the overall nematic order parameter $S$.  
The range of temperatures where coexistence between the nematic and an isotropic phase is observed is limited, but it
increases as the value of $\kappa'$ decreases. In the $NVT$ simulations 
we find smectics in equilibrium with a virtually empty vapour phase at
low temperatures, where the typical smectic layering emerges from the interface. The smectic ordering
can also been characterized by the evaluation of the parallel and perpendicular correlation distribution functions inside
the condensed phase layer, in a similar way as it is done in bulk. An analysis of the in-layer
structure shows hexatic order, indicating that the smectic phase is of $B$-type.
On the other hand, we do not find vapour-isotropic liquid coexistence, even 
for the 
smallest value of $\kappa'=0.15$. This implies that there is no critical point. As a consequence, the nematic
phase is in equilibrium with an isotropic fluid with a density with changes continuously from vapour-like to liquid-like 
values. Thus, as the temperature is increased, only a homogeneous isotropic fluid is observed in the $NVT$ simulations,
since the the system cannot develop interfacial configurations when the coexisting densities are below some threshold due
to the total number of particles constraint. 
So, in an effective way there is an upper limit for the temperature to observe coexistence between the nematic and
isotropic phase in our method.

\begin{figure}
\begin{center}
\includegraphics[width=0.8\columnwidth]{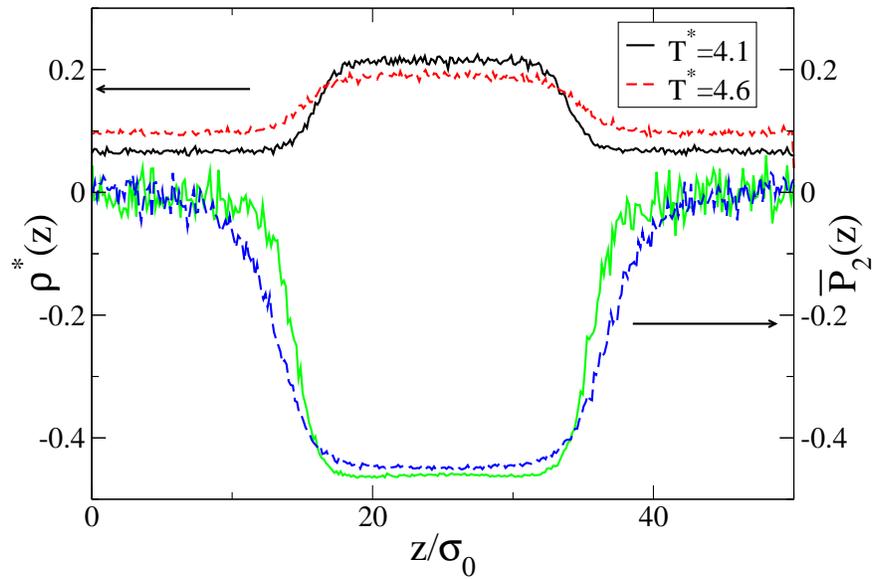}%
\caption{Density $\rho^*$ (upper curves) and orientational order $\overline{P}_2(z)$ (lower curves) profiles 
for a Gay-Berne fluid with $\kappa=4$ and $\kappa'=0.15$, for $T^*=4.1$ (continuous lines) and 
$T^*=4.6$ (dashed lines). 
\label{fig1}}
\end{center}
\end{figure}

\begin{figure}
\begin{center}
\includegraphics[width=0.8\columnwidth]{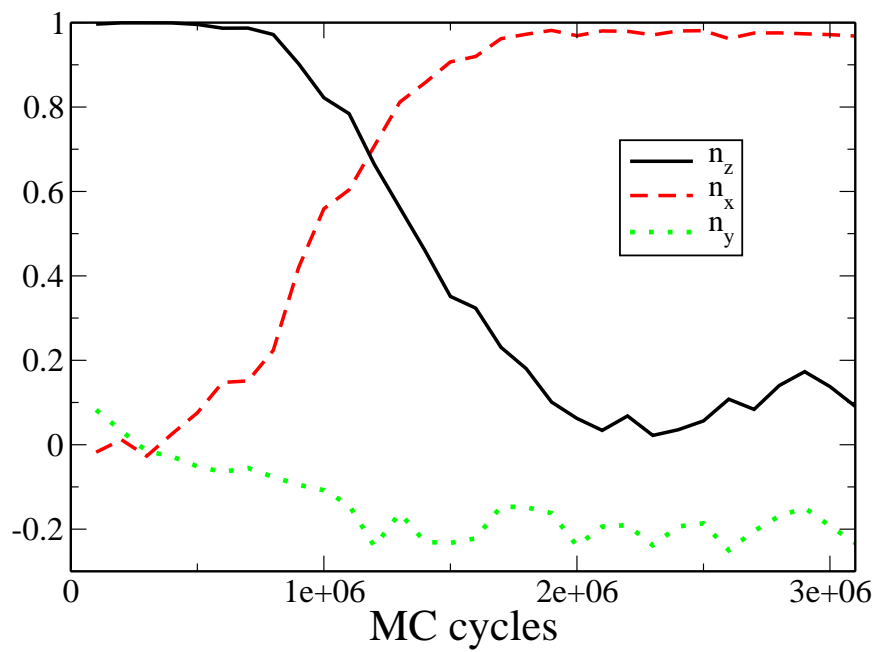}%
\caption{Evolution of the nematic director components for a Gay-Berne fluid with $\kappa=4$ and $\kappa'=1$ for 
$T^*=0.8$ in presence of a nematic-vapour interface, starting at a homeotropic configuration. 
\label{fig2}}
\end{center}
\end{figure}

Fig. \ref{fig1} shows the density and $\overline{P}_2$ profiles for $\kappa =4$, $\kappa'=0.15$ and $T^*=4.1$ and $T^*=4.6$. None of these profiles show any structure on the interfaces, but there is monotonous crossover from the bulk 
isotropic to the bulk nematic values. 
Note that at the highest temperature the isotropic fluid density $\rho^*=0.101$ is close to the corresponding to the
nematic case $\rho^*=0.187$. In both cases, the nematic phase shows a planar anchoring on the nematic-isotropic fluid 
interface, since $\overline{P}_2$ is negative in the nematic phase.
For prolate particles, we only observe planar anchoring for all the cases we studied (see Fig. 
\ref{fig0}a). 
In order to confirm this results, we performed additional simulations for selected conditions from initial conditions 
where the nematic showed a homeotropic
alignment. Fig. \ref{fig2} shows that the different nematic director components in the nematic region for $\kappa=4$,
$\kappa'=1$ and $T^*=0.80$ at nematic-vapour coexistence evolve to the planar configuration after approximately $10^6$ 
cycles.

Oblate molecules (i.e. $\kappa<1$) show a richer phenomenology. We choose two values of the molecular elongation
parameter $\kappa =0.3$ and $0.5$. The results obtained from our computer simulations are summarised in Table \ref{table4}.
Now we find coexistence between the vapour and both nematic and isotropic liquid phases. Again the temperature range
where nematic-vapour coexistence appears is limited, being slightly larger for $\kappa=0.3$ than for $\kappa=0.5$. 
At low temperatures, columnar phases appear in equilibrium with 
the vapour. On the other hand, as the temperature raises, a vapour-isotropic liquid-nematic triple point appears, and above
it the vapour is in equilibrium with an isotropic liquid. As in the case of the smectic phase for prolate
molecules, columnar phase can be identify from the analysis of the parallel and perpendicular pair correlation functions in
the condensed phase layer.

\begin{figure}
\begin{center}
\includegraphics[width=0.8\columnwidth]{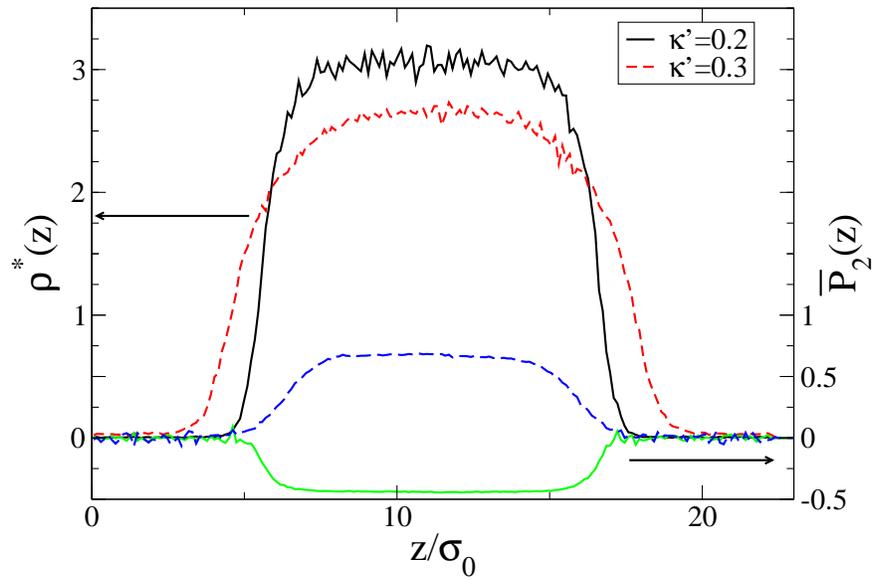}%
\caption{Density $\rho^*$ (upper curves) and orientational order $\overline{P}_2(z)$ (lower curves) profiles
at $T^*=2.20$ for Gay-Berne fluids with $\kappa=0.3$ and $\kappa'=0.2$ (continuous lines) and $\kappa'=0.3$ 
(dashed lines). 
\label{fig3}}
\end{center}
\end{figure}

\begin{figure}
\begin{center}
\includegraphics[width=0.8\columnwidth]{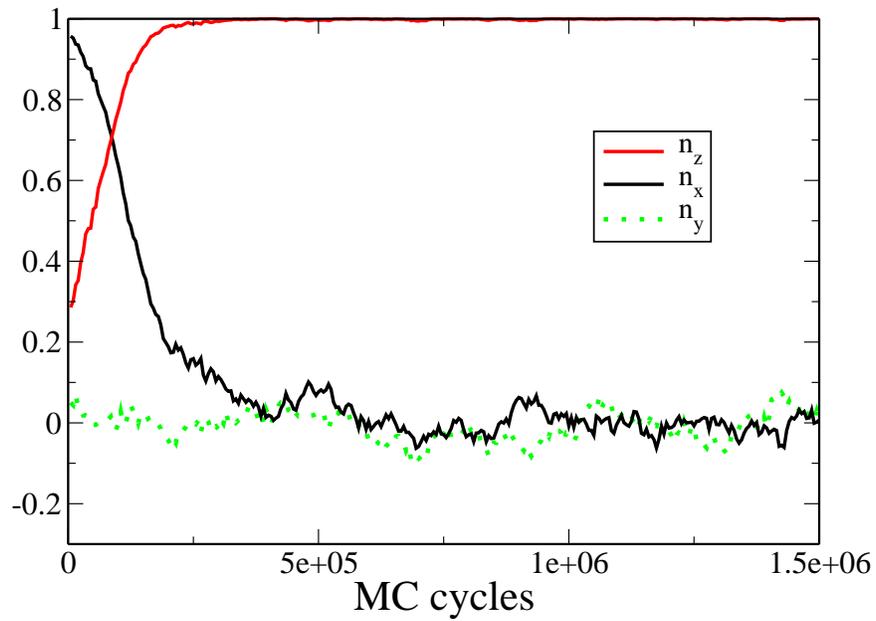}%
\caption{Evolution of the nematic director components for a Gay-Berne fluid with $\kappa=0.3$ and $\kappa'=1$ for 
$T^*=1.1$ in presence of a nematic-vapour interface, starting at a planar configuration. 
\label{fig4}}
\end{center}
\end{figure}

Fig. \ref{fig3} shows the density and $\overline{P}_2$ profiles for $\kappa =0.3$ and $T^*=2.20$ for $\kappa'=0.2$ and 
$\kappa'=0.3$. As for the prolate molecules case, these profiles smoothly crossover from vapour to nematic values across the nematic-vapour interface. We observe that for the smallest value of $\kappa'$ planar anchoring is observed 
since $\overline{P}_2$ is negative in the nematic layer, while for
the largest value the anchoring is homeotropic, i.e. $\overline{P}_2\approx S$ in the nematic layer, with $S$
being the nematic order parameter of the bulk phase. 
In fact, we observe that the anchoring is planar if $\kappa>\kappa'$, and homeotropic otherwise (see Table 
\ref{table4} and also Figs. \ref{fig0}b and \ref{fig0}c), independently of the temperature. 
This result is in agreement
with the arguments reported in Ref. \cite{Mills}. As in the case of the prolate molecules, we check this prediction by performing
additional simulations starting from initial conditions with the ``wrong'' nematic anchoring. Fig. \ref{fig4} shows the
evolution of the components of the global nematic director for $\kappa=0.3$, $\kappa'=1$ and $T^*=1.10$ at nematic-vapour
coexistence. We see that, starting from a planar 
anchoring, the nematic starts to reorient until it reaches the true homeotropic anchoring. We note that the number of
Monte Carlo cycles required is smaller than the corresponding value for prolate molecules. This may be due to the role
that the elastic deformations play in this orientation change.   

However, under some conditions a nematic-like layer is formed in the 
isotropic liquid-vapour interface (see Fig. \ref{fig0}d). 
Fig. \ref{fig4} shows the orientational order profiles
for $\kappa=0.5$ and $\kappa'=0.3$, which corresponds to planar anchoring, 
and $\kappa=0.5$ and $\kappa'=1$, which corresponds to homeotropic anchoring,
for a range of temperatures around the corresponding nematic-vapour-isotropic
triple point. In both cases, the density profiles (not shown) do not differ from
those discussed previously. However, in these cases we observe an enhancement
of the orientational ordering on the interfaces: molecules align preferentially
homeotropically on a layer at the interface, 
regardless the nematic anchoring away from the interface. This layer persists
even above the nematic-vapour-isotropic triple point. For the case 
$\kappa=0.5$ and $\kappa'=0.3$, the nematic-like layer is of molecular width
and it remains almost unchanged when crossing the triple-point temperature.
On the other hand, the nematic-like layer becomes very wide for $\kappa=0.5$
and $\kappa'=1$ above the triple point, and it is preceded by a small order 
deplection layer. In fact, we observe a nematic-vapour
interfacial configuration for $T^*=0.51$, which is above the triple
point from the $NpT$ simulations. Fig. \ref{fig6} shows the dependence 
on the number of particles of the orientational order profile of the 
nematic-vapour interfacial state for $\kappa=0.5$, $\kappa'=1.$ and 
$T^*=0.51$. We confirm that the isotropic phase is the equilibrium one at these
conditions ($\overline{P}_2\approx 0.1$ in the midpoint of the liquid layer), but its 
value decreases slowly with $N$. The explanation
of this phenomenon lies on the wetting properties of the isotropic-vapour 
interface, since it is likely to be wet by the nematic phase at the triple 
point. However, we have not done a systematic study to confirm this point. 
In any case, the existence of wetting implies that the simulations must be 
done more carefully, as there is an additional finite-size dependence.

The values of the reduced nematic-vapour surface tensions reported in Tables \ref{table3} and \ref{table4} show a strong
dependence on the Gay-Berne potential parameters (see also insets of Figs. 
\ref{fig7} and \ref{fig8}). 
However, a scaling behaviour is observed with respect to the potential
parameters by considering the arguments outlined in Ref. \cite{Mills}. So, for homeotropic anchoring, the surface tension
is expected to scale as ${N^H} \epsilon_{ee}/A$, where $N^H$ is the number of particles on the interface of area
$A$, an $\epsilon_{ee}$ is the scale of the intermolecular potential energy in the end-to-end configuration. Since
${N^H}/{A} \sim {1}/{\pi \sigma_{ss}^2}$, where $\sigma_{ss}$ is the closest distance between two particles in
the side-to-side configuration, we find that $\gamma^H\sim \epsilon_{ee} /{\sigma_{ss}^2}$. 
For planar anchoring, similar arguments yield $\gamma^P \sim N^P \epsilon_{ss}/A$, and $N^P/A\sim 1/\pi 
\sigma_{ss}\sigma_{ee}$, so $\gamma^P \sim \epsilon_{ss}/\sigma_{ss}\sigma_{ee}$. Thus, $\gamma^H/\gamma^P\sim (\epsilon_{ee} \sigma_{ee})/(\epsilon_{ss}
\sigma_{ss})=
\kappa/\kappa'$, which implies that homeotropic anchoring is favoured if $\kappa<\kappa'$, and planar anchoring otherwise.
Our results confirm this prediction, since homeotropic anchoring is only observed for oblate molecules with $\kappa \le 
\kappa'$ (see Table \ref{table4}). These results suggest that we should
use rescaled surface tensions 
$\gamma'=\gamma \sigma_{ss}^2/\epsilon_{ee}=2\kappa\kappa'\gamma^*/
(1+\kappa^2)$ for homeotropic anchoring, and 
$\gamma'=\gamma \sigma_{ss}\sigma_{ee}/\epsilon_{ss}=2\kappa^2\gamma^*/
(1+\kappa^2)$ for planar anchoring, to compare surface tensions for different 
values of $\kappa$ and $\kappa'$.
However, these arguments do not provide 
the scale for the temperature dependence of the nematic-isotropic surface 
tension. We argue that, when there is planar anchoring, the temperature scale
for the surface tension change is given by the largest in-plane intermolecular
attraction, which is $\epsilon_{ss}=\epsilon_0 (\kappa^2+1)/(2\kappa\kappa')$
if $\kappa'<1$, or $\epsilon_{ee}=\epsilon_0(\kappa^2+1)/(2\kappa)$ otherwise.
For homeotropic anchoring, this scale is given by $\epsilon_{ss}$. So,
we should use a rescaled temperature $T'=2\kappa\kappa' T^*/(1+\kappa^2)$
for cases with planar anchoring and $\kappa'<1$, and 
$T'=2\kappa T^*/(1+\kappa^2)$ otherwise.
Fig. \ref{fig7} represents the rescaled values of the nematic-isotropic
surface tensions as a function of the rescaled temperature for planar anchoring
and $\kappa'<1$. Note that under the rescaling the original curves, depicted
in the inset, collapse in a reasonable way to a master curve. Note that 
these curves were obtained for both prolate and oblate molecules with very 
different molecular characteristics. On the other hand, Fig. \ref{fig8}
plots the rescaled nematic-isotropic surface tensions vs. the rescaled 
temperature for homeotropic anchoring. The collapse is less clear than 
for the planar anchoring case, but still we can see that trends are common
except maybe for the case $\kappa=0.3$ and $\kappa'=1$, which has a 
larger slope than the other curves. In any case, by comparison with the 
original data shown in the inset, the rescaling has a reasonable performance.

\section{Conclusions \label{sec4}}

In this paper we present Monte Carlo simulations of the nematic-vapour 
interface for the Gay-Berne model, corresponding to either prolate and oblate
molecules. We characterize the density and orientational order parameter 
profiles, as well as the anchoring of the nematic phase with respect to
the nematic-vapour interface. For $\kappa>\kappa'$ we find that the anchoring
is planar, and homeotropic otherwise, regardless the particles being prolate
or oblate. These findings are in agreement with heuristic arguments reported
in the literature \cite{Mills}. Under some circumstances, an enhanced 
orientational order is observed in the nematic-vapour interface. 
On the other hand, we studied the 
nematic-vapour surface tension as a function of the temperature. By a 
generalization of the arguments presented in Ref. \cite{Mills}, we 
find that by rescaling the surface tension and the temperature by factors which
depend on both $\kappa$ and $\kappa'$, our simulation data satisfy 
approximately corresponding states laws for planar and homeotropic anchoring.
This is a remarkable result, since they work for very different molecular 
geometries. We expect that these findings can be corroborated with other 
liquid crystal models.

\section*{Acknowledgments}
We acknowledge financial support from the Portuguese Foundation for Science and Technology under Contract No. EXCL/FIS-NAN/0083/2012, the Spanish Ministerio de Econom\'{\i}a y Competitividad through
grant no. FIS2012-32455, and Junta de Andaluc\'{\i}a through grant no. P09-FQM-4938, all co-funded
by the EU FEDER.

\clearpage

\begin{table}[t] 
\tbl{Influence of the Gay-Berne potential cut-off on the surface tension.\label{table1}} 
{\begin{tabular}{ccccc}\toprule
$\kappa$ & $\kappa'$ & $ T^* $ & $r_c^*$ & $\gamma$  \\
\hline
3 & 1 & 0.55 & 4 & 0.416  \\
3 & 1 & 0.55 & 5 & 0.470  \\
3 & 1 & 0.55 & 7 & 0.471  \\
3 & 1 & 0.59 & 4 & 0.369  \\
3 & 1 & 0.59 & 5 & 0.415  \\
3 & 1 & 0.59 & 7 & 0.412  \\
\bottomrule
\end{tabular}}
\end{table}

\clearpage

\begin{table}[t]
\tbl{Comparison between the surface tension values $\gamma_{EE}$ obtained with different expanded ensemble techniques
for $\kappa=6$ and number of particles $N$. $p_1$ is the probability to visit the
state $1$ (see text for explanation).\label{table2}}
{\begin{tabular}{cccc}\toprule
Changed box side & $ N $ & $p_1$ & $\gamma_{EE}$ \\
\hline
$L_x$ & 1372 & 0.490 (22) & 0.479 (17) \\
       & 4000 & 0.502 (21) & 0.379 (12) \\
$L_y$ & 1372 & 0.514 (30) & 0.395 (6)  \\
       & 4000 & 0.507 (26) & 0.374 (4)  \\
$L_x, L_x$ & 1372 & 0.506 (24) & 0.393 (9) \\
       & 4000 & 0.498 (17) & 0.378 (7) \\
\bottomrule
\end{tabular}}
\end{table}

\clearpage

\begin{table}[t]
\tbl{Summary of the simulation results for prolate molecules 
(see text for explanation).\label{table3}}
{\begin{tabular}{ccccccccc}\toprule
$\kappa$ & $\kappa'$ & $T^*$ & $\rho_{bulk}^*$ & $Phase $ & $S$ & $\rho_{iso}^*$ & $\gamma_{virial}^*$ & $Anchoring$\\
\hline
4 & 1 & 0.55 & 0.283 (2) & $Sm$ & 0.967 (2) &       &            &     \\
4 & 1 & 0.60 & 0.261 (2) & $N$  & 0.954 (2) &       & 0.529 (15) & $P$ \\
4 & 1 & 0.65 & 0.251 (2) & $N$  & 0.937 (2) & 0.001 & 0.445 (14) & $P$ \\
4 & 1 & 0.70 & 0.241 (1) & $N$  & 0.923 (3) & 0.005 & 0.417 (13) & $P$ \\
4 & 1 & 0.75 & 0.233 (2) & $N$  & 0.900 (3) & 0.008 & 0.314 (10) & $P$ \\
4 & 1 & 0.80 & 0.221 (2) & $N$  & 0.871 (6) & 0.015 & 0.209 (10) & $P$ \\
4 & 0.5 & 0.80 & 0.297 (2) & $Sm$ & 0.981 (1) &       &            &     \\
4 & 0.5 & 0.90 & 0.275 (2) & $N$ & 0.972 (2)  &       & 0.954 (30) & $P$ \\
4 & 0.5 & 1.00 & 0.264 (2) & $N$ & 0.962 (2)  &       & 0.822 (20) & $P$ \\
4 & 0.5 & 1.10 & 0.257 (2) & $N$ & 0.951 (2)  & 0.001 & 0.695 (18) & $P$ \\
4 & 0.5 & 1.20 & 0.243 (2) & $N$ & 0.946 (2)  & 0.005 & 0.522 (17) & $P$ \\
4 & 0.5 & 1.25 & 0.236 (2) & $N$ & 0.938 (2)  & 0.009 & 0.446 (15) & $P$ \\
4 & 0.5 & 1.30 & 0.229 (2) & $N$ & 0.929 (3)  & 0.017 (1) & 0.359 (12) & $P$ \\
4 & 0.5 & 1.35 & 0.220 (2) & $N$ & 0.909 (5)  & 0.032 (2) & 0.262 (7)  & $P$ \\
4 & 0.5 & 1.40 & 0.208 (4) & $N$ & 0.885 (6)  & 0.058 (2) & 0.185 (9)  & $P$ \\
4 & 0.25 & 2.00 & 0.274 (9) & $Sm$ & 0.993 (10) &     &            &     \\
4 & 0.25 & 2.10 & 0.249 (10) & $N$ & 0.961 (18) & 0.006  & 0.921 (46) & $P$ \\
4 & 0.25 & 2.20 & 0.242 (10) & $N$ & 0.957 (19) & 0.011  & 0.853 (56) & $P$ \\
4 & 0.25 & 2.30 & 0.235 (10) & $N$ & 0.952 (20) & 0.023 (1) & 0.654 (50) & $P$ \\
4 & 0.25 & 2.40 & 0.227 (10) & $N$ & 0.940 (17) & 0.035 (1) & 0.506 (42) & $P$ \\
4 & 0.25 & 2.50 & 0.215 (9)  & $N$ & 0.931 (12) & 0.055 (1) & 0.387 (41) & $P$ \\
4 & 0.25 & 2.60 & 0.203 (6)  & $N$ & 0.903 (22) & 0.072 (2) & 0.303 (33) & $P$ \\
4 & 0.25 & 2.70 & 0.201 (9)  & $N$ & 0.875 (26) & 0.086 (3) & 0.243 (10) & $P$ \\
4 & 0.25 & 2.80 & 0.183 (9)  & $N$ & 0.761 (42) & 0.095 (4) & 0.197 (11) & $P$ \\
4 & 0.25 & 2.90 & 0.174 (7)  & $N$ & 0.703 (52) & 0.104 (7) & 0.147 (10) & $P$ \\
4 & 0.15 & 3.40 & 0.246 (1)  & $Sm$ & 0.96 (1) &       &            & \\
4 & 0.15 & 3.50 & 0.242 (1)  & $N$ & 0.95 (1) & 0.015  & 1.053 (80) & $P$ \\
4 & 0.15 & 3.60 & 0.237 (1)  & $N$ & 0.95 (1) & 0.025  & 1.005 (78) & $P$ \\
4 & 0.15 & 3.70 & 0.231 (1)  & $N$ & 0.94 (1) & 0.032  & 0.953 (69) & $P$ \\
4 & 0.15 & 3.80 & 0.226 (1)  & $N$ & 0.94 (2) & 0.042 (1) & 0.856 (62) & $P$ \\
4 & 0.15 & 3.90 & 0.218 (1)  & $N$ & 0.93 (2) & 0.055 (1) & 0.814 (65) & $P$ \\
4 & 0.15 & 4.00 & 0.209 (1)  & $N$ & 0.93 (2) & 0.062 (1) & 0.621 (53) & $P$ \\
4 & 0.15 & 4.10 & 0.215 (1)  & $N$ & 0.90 (3) & 0.071 (2) & 0.502 (25) & $P$ \\
4 & 0.15 & 4.20 & 0.205 (2)  & $N$ & 0.85 (4) & 0.082 (2) & 0.423 (22) & $P$ \\
4 & 0.15 & 4.30 & 0.202 (4)  & $N$ & 0.84 (5) & 0.085 (4) & 0.356 (15) & $P$ \\
4 & 0.15 & 4.40 & 0.195 (3)  & $N$ & 0.71 (6) & 0.091 (5) & 0.337 (16) & $P$ \\
4 & 0.15 & 4.50 & 0.190 (2)  & $N$ & 0.68 (7) & 0.095 (5) & 0.282 (15) & $P$ \\
4 & 0.15 & 4.60 & 0.187 (5)  & $N$ & 0.62 (7) & 0.101 (5) & 0.250 (18) & $P$ \\
4 & 0.15 & 4.70 & 0.184 (5)  & $N$ & 0.63 (8) & 0.112 (7) & 0.227 (18) & $P$ \\
6 & 0.5 & 1.50 & 0.187 (3) & $Sm$ &  0.978 (3) &       &            &     \\
6 & 0.5 & 1.60 & 0.178 (6)  & $N$ &  0.963 (3) & 0.005 & 0.678 (25) & $P$ \\
6 & 0.5 & 1.70 & 0.172 (10) & $N$ &  0.905 (4) & 0.012 & 0.502 (20) & $P$ \\
6 & 0.5 & 1.75 & 0.168 (11) & $N$ &  0.873 (5) & 0.017 & 0.445 (22) & $P$ \\
6 & 0.5 & 1.80 & 0.161 (10) & $N$ &  0.817 (7) & 0.021 & 0.385 (21) & $P$ \\
6 & 0.5 & 1.85 & 0.157 (9)  & $N$ &  0.797 (9) & 0.025 & 0.361 (22) & $P$ \\
6 & 0.5 & 1.90 & 0.152 (11) & $N$ &  0.781 (11)& 0.031 & 0.324 (15) & $P$ \\
6 & 0.5 & 2.00 & 0.148 (10) & $N$ &  0.774 (10)& 0.042 & 0.220 (13) & $P$ \\
6 & 0.5 & 2.10 & 0.137 (7)  & $N$ &  0.743 (11)& 0.049 (1) & 0.169 (10) & $P$ \\
6 & 0.5 & 2.20 & 0.128 (10) & $N$ & 0.735 (12) & 0.055 (1) & 0.152 (10) & $P$ \\
6 & 0.5 & 2.30 & 0.126 (15) & $N$ & 0.740 (15) & 0.061 (1) & 0.127 (6)  & $P$ \\
6 & 0.25 & 3.30 & 0.185 (2) & $Sm$ & 0.984 (2) &       &            &     \\
6 & 0.25 & 3.40 & 0.163 (4) & $N$ & 0.935 (3)  & 0.027 & 0.605 (29) & $P$ \\
6 & 0.25 & 3.50 & 0.159 (5) & $N$ & 0.921 (3)  & 0.031 & 0.476 (26) & $P$ \\
6 & 0.25 & 3.60 & 0.153 (4) & $N$ & 0.902 (4)  & 0.036 & 0.435 (27) & $P$ \\
6 & 0.25 & 3.70 & 0.149 (2) & $N$ & 0.874 (7)  & 0.040 & 0.354 (41) & $P$ \\
6 & 0.25 & 3.80 & 0.145 (3) & $N$ & 0.831 (6)  & 0.044 & 0.318 (19) & $P$ \\
6 & 0.25 & 3.90 & 0.140 (4) & $N$ & 0.804 (7)  & 0.047 (1) & 0.301 (20) & $P$ \\
\bottomrule
\end{tabular}}
\end{table}

\clearpage

\begin{table}[t]
\tbl{Summary of the simulation results for oblate molecules (see text for
explanation).\label{table4}}
{\begin{tabular}{ccccccccc}
$\kappa$ & $\kappa'$ & $T^*$ & $\rho_{bulk}^*$ & $Phase $ & $S$ & $\rho_{vapor}^*$ & $\gamma_{virial}^*$ & $Anchoring$\\
\hline
0.3 & 0.4 & 1.20 & 3.41 (3)  & $C$ & 0.92 &       &               &     \\
0.3 & 0.4 & 1.30 & 3.24 (2)  & $N$ & 0.89 &       & 12.28 (12)  & $H$ \\
0.3 & 0.4 & 1.40 & 3.17 (2)  & $N$ & 0.87 &       & 11.36 (11)  & $H$ \\
0.3 & 0.4 & 1.50 & 3.12 (1)  & $N$ & 0.84 &       & 10.51 (10)  & $H$ \\
0.3 & 0.4 & 1.60 & 3.01 (1)  & $N$ & 0.81 &       &  9.25  (9) & $H$ \\
0.3 & 0.4 & 1.70 & 2.94 (1)  & $N$ & 0.77 &       &  8.20  (9) & $H$ \\
0.3 & 0.4 & 1.80 & 2.84 (1)  & $N$ & 0.73 & 0.01  &  7.06  (8) & $H$ \\
0.3 & 0.4 & 1.90 & 2.73 (1)  & $N$ & 0.65 & 0.02  &  5.73  (6) & $H$ \\
0.3 & 0.4 & 2.00 & 2.21 (1)  & $I$ &      & 0.04  &  3.82 (6)   &     \\
0.3 & 0.3 & 1.40 & 3.38 (7)  & $C$ & 0.95 &       &               &     \\
0.3 & 0.3 & 1.50 & 3.24 (3)  & $N$ & 0.88 &       & 14.00 (2)  & $H$ \\
0.3 & 0.3 & 1.60 & 3.16 (3)  & $N$ & 0.87 &       & 12.80 (2)  & $H$ \\
0.3 & 0.3 & 1.70 & 3.10 (4)  & $N$ & 0.84 &       & 11.81 (2)  & $H$ \\
0.3 & 0.3 & 1.80 & 3.03 (4)  & $N$ & 0.83 &       & 10.58 (1)  & $H$ \\
0.3 & 0.3 & 1.90 & 2.95 (4)  & $N$ & 0.78 &       &  9.43 (1)  & $H$ \\
0.3 & 0.3 & 2.00 & 2.88 (5)  & $N$ & 0.73 & 0.01  &  8.30 (2)  & $H$ \\
0.3 & 0.3 & 2.10 & 2.78 (6)  & $N$ & 0.66 & 0.02  &  6.96 (1)  & $H$ \\
0.3 & 0.3 & 2.20 & 2.65 (7)  & $N$ & 0.55 & 0.03  &  5.56 (1)  & $H$ \\
0.3 & 0.3 & 2.30 & 2.08 (8)  & $I$ &      & 0.05  &  3.51 (1)  &     \\
0.3 & 0.2 & 2.10 & 3.14 (10) & $C$ & 0.93 &       &            &     \\
0.3 & 0.2 & 2.20 & 3.06 (10) & $N$ & 0.90 &       & 12.83 (16) & $P$ \\
0.3 & 0.2 & 2.30 & 3.00 (15) & $N$ & 0.89 &       & 11.52 (15) & $P$ \\
0.3 & 0.2 & 2.40 & 2.94 (14) & $N$ & 0.79 & 0.01  & 10.31 (15) & $P$ \\
0.3 & 0.2 & 2.50 & 2.86 (10) & $N$ & 0.76 & 0.02  &  8.99 (13) & $P$ \\
0.3 & 0.2 & 2.60 & 2.79 (10) & $N$ & 0.72 & 0.03  &  7.78 (13) & $P$ \\
0.3 & 0.2 & 2.70 & 2.70 (9)  & $N$ & 0.65 & 0.05  &  6.37 (10) & $P$ \\
0.3 & 0.2 & 2.80 & 2.58 (9)  & $N$ & 0.47 & 0.06  &  4.43 (9)  & $P$ \\
0.3 & 0.2 & 2.90 & 2.08 (10) & $I$ &      & 0.10  &  2.82 (13) &     \\
0.5 & 1.0 & 0.30 & 2.12 (1)   & $C$ & 0.85  &       &               &     \\
0.5 & 1.0 & 0.40 & 1.97 (2)   & $N$ & 0.77  &       &  4.22 (8)   & $H$ \\
0.5 & 1.0 & 0.50 & 1.86 (2)   & $N$ & 0.50  &       &  3.56 (8)   & $H$ \\
0.5 & 1.0 & 0.60 & 1.75 (2)   & $I$ &       & 0.001 &  3.27 (9)   &     \\
0.5 & 0.7 & 0.30 & 2.12 (10)  & $C$ & 0.83  &       &               &     \\
0.5 & 0.7 & 0.40 & 1.99 (9)   & $N$ & 0.74  &       &  5.16 (9)   & $H$ \\
0.5 & 0.7 & 0.50 & 1.89 (6)   & $N$ & 0.48  &       &  4.28 (6)   & $H$ \\
0.5 & 0.7 & 0.60 & 1.78 (13)  & $I$ & 0.45  &       &  3.94 (5)   &     \\
0.5 & 0.6 & 0.30 & 2.15 (9)   & $C$ & 0.85  &       &               &     \\
0.5 & 0.6 & 0.40 & 2.03 (5)   & $N$ & 0.73  &       &  5.31 (9)   & $H$ \\
0.5 & 0.6 & 0.50 & 1.94 (9)   & $N$ & 0.61  &       &  4.61 (7)   & $H$ \\
0.5 & 0.6 & 0.60 & 1.80 (7)   & $I$ &       &       &  4.18 (6)   &     \\
0.5 & 0.5 & 0.40 & 2.16 (3)   & $C$ & 0.84  &       &               &     \\
0.5 & 0.5 & 0.50 & 1.97 (4)   & $N$ & 0.62  &       &  5.31 (9)   & $H$ \\
0.5 & 0.5 & 0.60 & 1.82 (6)   & $I$ &       &       &  4.69 (8)   &     \\
0.5 & 0.4 & 0.50 & 2.04 (5)   & $C$ & 0.87  &       &               &     \\
0.5 & 0.4 & 0.60 & 1.91 (7)   & $N$ & 0.48  &       &  5.35 (8)   & $P$ \\
0.5 & 0.4 & 0.70 & 1.79 (6)   & $I$ &       &       &  4.64 (8)   &     \\
0.5 & 0.3 & 0.70 & 1.95 (4)   & $C$ & 0.80  &       &               &     \\
0.5 & 0.3 & 0.80 & 1.85 (5)   & $N$ & 0.44  &       &  5.24 (7)   & $P$ \\
0.5 & 0.3 & 0.90 & 1.73 (3)   & $I$ &       &       &  4.45 (7)   &     \\
0.5 & 0.2 & 1.00 & 2.04 (3)   & $C$ & 0.873 &       &               &     \\
0.5 & 0.2 & 1.10 & 1.85 (11)  & $N$ & 0.662 &       &  5.63 (7)   & $P$ \\
0.5 & 0.2 & 1.20 & 1.76 (12)  & $N$ & 0.536 &       &  4.98 (7)   & $P$ \\
0.5 & 0.2 & 1.30 & 1.64 (15)  & $I$ &       &       &  4.59 (4)   &     \\
\bottomrule
\end{tabular}}
\end{table}
 
\clearpage

\clearpage

\clearpage

\clearpage

\clearpage

\clearpage

\begin{figure}
\begin{center}
\includegraphics[width=0.8\columnwidth]{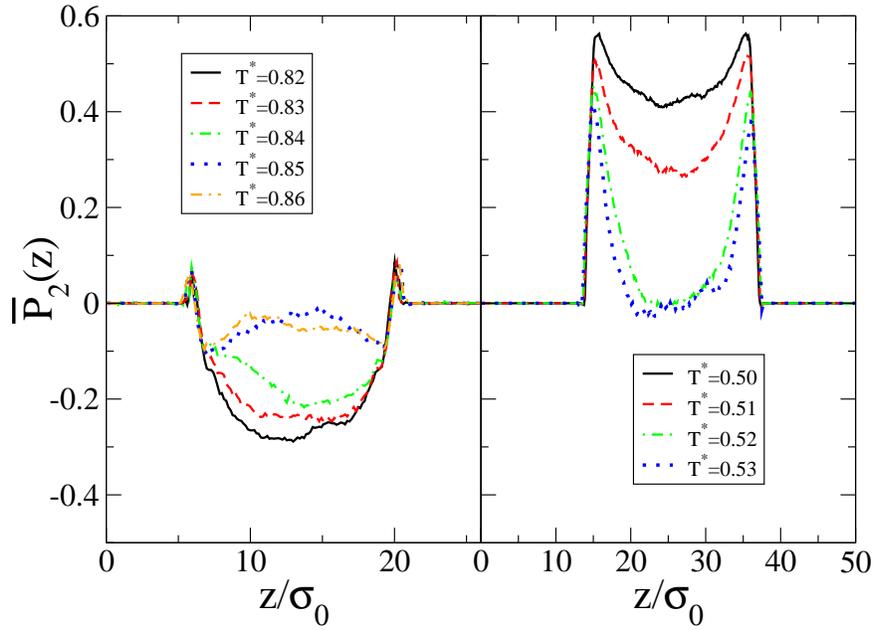}%
\caption{Left panel: plot of the orientational order profile for 
$\kappa=0.5$, $\kappa'=0.3$, $N=4000$ and 
$T^*=0.82$ (continuous line), $0.83$ (dashed line),
$0.84$ (dot-dashed line), $0.85$ (dotted line) and $0.86$ (dot-double dashed line). The nematic-isotropic-vapour triple point temperature 
is $T^*=0.845\pm0.005$.
Right panel: plot of the orientational order profile for 
$\kappa=0.5$, $\kappa'=1.0$, $N=4116$ and 
$T^*=0.50$ (continuous line), $0.51$ (dashed line),
$0.52$ (dot-dashed line) and $0.53$ (dotted line). 
The nematic-isotropic-vapour triple point temperature is $T^*=0.505\pm0.005$.
\label{fig5}}
\end{center}
\end{figure}

\clearpage

\begin{figure}
\begin{center}
\includegraphics[width=0.8\columnwidth]{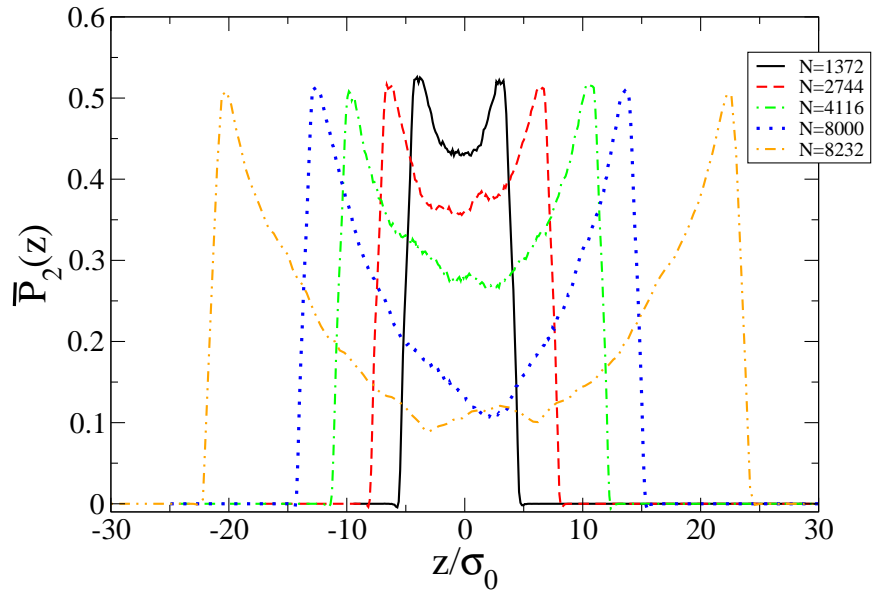}%
\caption{Dependence on the number of particles $N$ 
of the simulated orientational
order profile for $\kappa=0.5$, $\kappa'=1.0$ and $T^*=0.51$: 
$N=1372$ (continuous line), $2744$ (dashed line), $4116$ (dot-dashed line),
$8000$ (dotted line) and $8232$ (double dot-dashed line).
\label{fig6}}
\end{center}
\end{figure}

\clearpage

\begin{figure}
\begin{center}
\includegraphics[width=0.8\columnwidth]{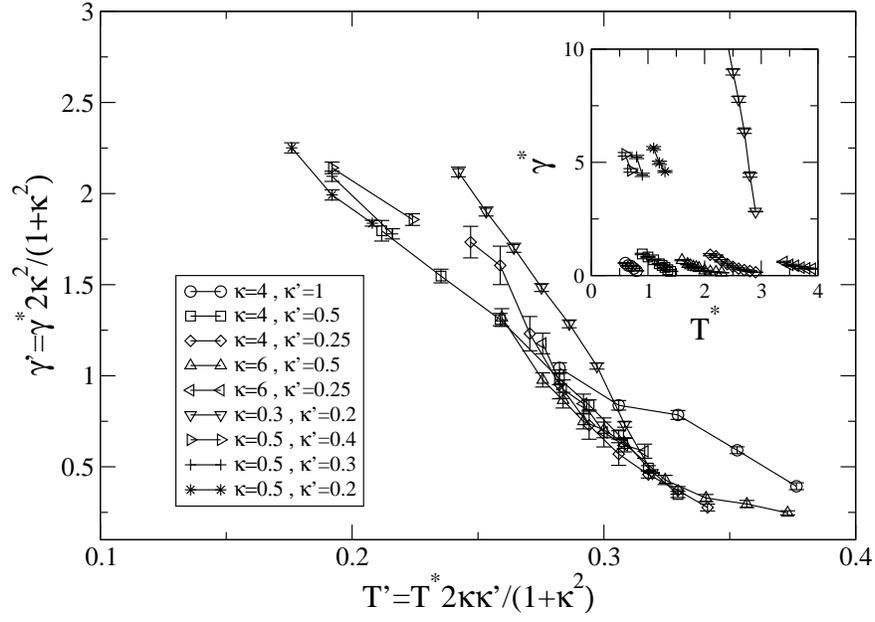}%
\caption{Plot of the rescaled nematic-vapour surface tension 
$\gamma'=2\kappa^2\gamma^*/(1+\kappa^2)$ as a function of the rescaled 
temperature $T'=2\kappa\kappa'T^*/(1+\kappa^2)$  for
planar anchoring: $\kappa=4$, $\kappa'=1$ (circles); $\kappa=4$, $\kappa'=0.5$
(squares);  $\kappa=4$, $\kappa'=0.25$ (diamonds); $\kappa=6$, $\kappa'=0.5$
(triangles up); $\kappa=6$, $\kappa'=0.25$ (triangles left);  $\kappa=0.3$, 
$\kappa'=0.2$ (triangles down); $\kappa=0.5$, $\kappa'=0.4$ (triangles right);
$\kappa=0.5$, $\kappa'=0.3$ (pluses) and $\kappa=0.5$, $\kappa'=0.2$ (stars).
Inset: Plot of the reduced $\gamma^*$ vs $T^*$. The meaning of the symbols is 
the same as in the main plot.
\label{fig7}}
\end{center}
\end{figure}

\clearpage

\begin{figure}
\begin{center}
\includegraphics[width=0.8\columnwidth]{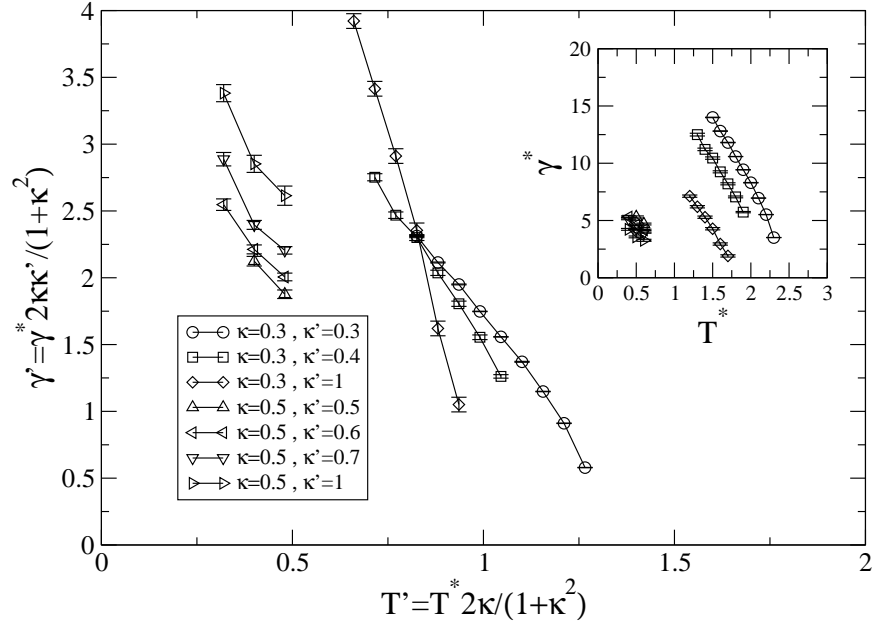}%
\caption{Plot of the rescaled nematic-vapour surface tension 
$\gamma'=2\kappa\kappa'\gamma^*/(1+\kappa^2)$ as a function of the rescaled 
temperature $T'=2\kappa T^*/(1+\kappa^2)$  for
homeotropic anchoring: $\kappa=0.3$, $\kappa'=0.3$ (circles); $\kappa=0.3$, $\kappa'=0.4$
(squares);  $\kappa=0.3$, $\kappa'=1$ (diamonds); $\kappa=0.5$, $\kappa'=0.5$
(triangles up); $\kappa=0.5$, $\kappa'=0.6$ (triangles left);  $\kappa=0.5$, 
$\kappa'=0.7$ (triangles down) and $\kappa=0.5$, $\kappa'=1$ (triangles right).
Inset: Plot of the reduced $\gamma^*$ vs $T^*$. The meaning of the symbols is 
the same as in the main plot.
\label{fig8}}
\end{center}
\end{figure}
\end{document}